\documentclass[aps,twocolumn]{revtex4-1}
\usepackage{epsfig}
\usepackage{textcomp}
\usepackage{float}
\usepackage{color}
\usepackage{graphicx}

\begin{document}

\title{Furrows in the wake of propagating d-cones}
\author{Omer Gottesman$^{1,2}$, Efi Efrati$^{3}$, Shmuel Rubinstein$^{1,2}$}

\affiliation{$^{1}$Department of Physics of Complex Systems, Weizmann Institute of Science, Rehovot 76100, Israel}
\affiliation{$^{2}$School of Engineering and Applied Sciences, Harvard University, Cambridge, Massachusetts 02138, USA}
\affiliation{$^{3}$James Franck Institute, The University of Chicago. 929 E. 57 st, Chicago, IL 60637, USA}

\begin{abstract}

We investigate the formation dynamics of plastic creases in thin elasto-plastic sheets. In contrast to the commonly accepted description of crumpled thin sheets, which asserts that creases form only by elastic interaction between \emph{two} d-cones, the creases we study in this letter are created by plastic deformations left in the wake of a \emph{single} propagating d-cone. Upon application of load, a d-cone initially remains stationary and responds by deforming globally. However, above a critical load, the d-cone undergoes a sharpening transition that focuses the stresses at its tip, allowing it to propagate along the sheet, leaving a furrow-like scar in its wake. Our results show that the dynamics of plastic defect creation are important for predicting the final geometry and statistics of a defect network in a crumpled thin sheet.

\end{abstract}
\maketitle

When handling a large sheet of paper one must take special care not to impose boundary conditions that will result in local strains that yield the material and leave permanent defects. In fact, keeping a large sheet of paper smooth is a difficult task; more often than not wrinkles and scars appear on the surface of the paper. An example for this may be found in the mesmerizing network of scars which decorate the face of a sheet of paper after it has been crumpled to the shape of a ball and then opened \cite{witten2007stress}. In recent years, crumpled paper has indeed attracted much attention in the scientific community and has been investigated analytically \cite{amar1997crumpled,lobkovsky1995scaling}, numerically \cite{vliegenthart2006forced,tallinen2008effect}, and experimentally \cite{cerda1999conical,cambou2011three,deboeuf2013comparative,boudaoud2000dynamics}.

In thin elastic sheets, bending deformations are energetically favorable compared to stretching deformations. Thus, a thin elastic sheet will preferentially bend to accommodate its prescribed boundary conditions and whenever possible will exhibit very little if any in plane stretching. In many cases, however, the prescribed boundary conditions allow no such stretch-free configuration with finite bending energy; such is the case when a thin sheet is confined to a small volume \cite{venkataramani2000limitations}. In this case the unavoidable in-plane stretching localizes to point-like and line-like objects termed d-cones and stretching ridges respectively \cite{witten2007stress,lobkovsky1995scaling,cerda2005confined,chaieb1998experimental}. It is believed that the crumpled configuration of a highly confined thin sheet can be fully described using only these two deformation modes. To calculate the geometry of these singular deformations, the energetic cost of in-plane stretching is balanced against the otherwise diverging bending energy. This approach has been very successful and accurately predicts the spatial structure of an isolated d-cone \cite{cerda2005confined,cerda1998conical} and of a single stretching ridge \cite{lobkovsky1995scaling}.

\begin{figure}
\includegraphics[width=3.5 in,clip=]{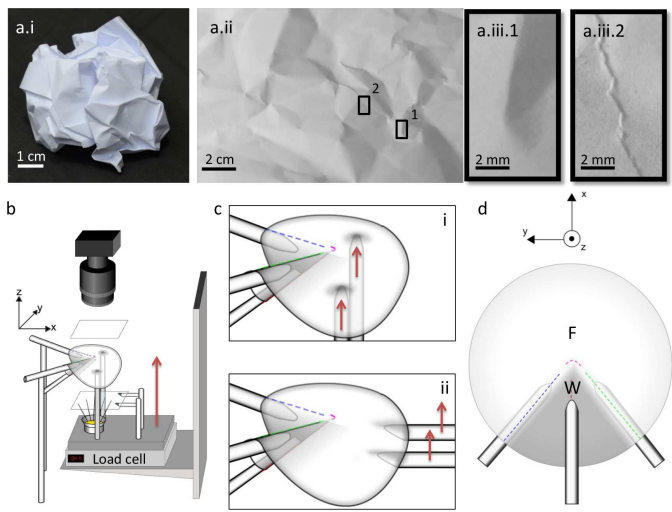}
\caption{\textbf{Experimental setup.} (a) A ball of crumpled paper (i) and the crease pattern on the opened sheet (ii) exhibit the classical straight ridges connecting two d-cones (iii.1) as well as ragged scars created by a single propagating d-cone (iii.2). (b) A schematic illustration of the experimental setup. (c) An illustration of the two loading geometries used: vertical (i) and horizontal (ii). (d) A schematic top view of the supporting posts and the sheet, indicating the front (F) and the wake (W) of the advancing d-cone.}
\label{fig1}
\end{figure}

The purely elastic considerations used to explain d-cone and ridge structures overlook the important role that the time evolution of plastic defect creation has in determining the final geometry and statistics of the network \cite{tallinen2008effect}. One successful recent approach incorporates plasticity through the mechanism of hierarchical ridge breaking \cite{blair2005geometry,wood2002witten}. In this framework, existing defects are considered immobile and the force resisting further crumpling is associated with the nucleation of new point-defects and ridges \cite{houle1996acoustic,matan2002crumpling}. In contrast, for purely elastic sheets, point defects and their connecting ridges respond to loading by the redistribution of the elastic energy trough defect migration \cite{aharoni2010direct}, analogues to the role played by dislocations in an ordered solid. However, most materials are neither purely elastic nor fully plastic. An elasto-plastic material will not only nucleate new defects in response to loading but will also allow preexisting defects to move, leaving scars in their wake; thus, the crumpling of elasto-plastic material is expected to exhibit highly non-trivial dissipative dynamics. The critical role that dynamics and plasticity play in crumpling can be observed by opening up a crumpled sheet of regular printing paper and closely examining its creases. Apart from the straight and relatively blunt ridges connecting at vertices, the crumpled paper also exhibits sharper, more crooked and ragged scars, as shown in Fig. \ref{fig1}(a). These scars, which are reminiscent of furrows in a plowed terrain, result from the propagation of single d-cones across the paper.

In this letter we introduce and characterize a new type of crease that is inherently plastic and forms by the propagation of a single point-defect. We strain a preexisting d-cone in an elsto-plastic thin sheet and show it remains pinned up to a critical loading force, $F_y$, which scales quadratically with the thickness of the thin sheet. When $F_y$ is reached, the singular structure at the apex of the d-cone sharpens abruptly. The resulting focusing of strains yields the material just ahead of the d-cone, allowing it to propagate, leaving a furrow-like scar in its wake.

To experimentally probe the propagation of point defects and resulting scarring in thin sheets we investigate single d-cones in commercial Mylar sheets. The sheets, $50\mu m$ to $250\mu m \pm 5\mu m$ in thickness, are cut to circles of 45 mm in radius. To controllably form an initial d-cone, we follow the procedure described by Cerda et al. \cite{boudaoud2000dynamics} and shown in supplementary movie S1. The Mylar sheet is manually pushed into a plastic ring-shaped constriction with an inner radius of 29 mm. The pushing force is applied to the center of the Mylar sheet perpendicular to the plane of the plastic ring and the total penetration depth of the center of the sheet into the plastic ring is 7.5mm. This procedure results in the formation of a single d-cone at the center of the Mylar sheet: a finite region supported on the plastic ring and forming a section of a cone and a free standing region concave at its center and flanked by two convex regions. The sheet is then removed from the confining ring while retaining the d-cone structure, as shown in Fig. \ref{fig1}(b) and \ref{fig1}(c), and transferred to a three point support, one below the sheet in the concave region, and two above the sheet in the convex regions. The geometry of the three point support, composed of 0.5" metal posts, is then fixed such that the central supporting post lies in the $xz$ plane at an angle of $\sim50^o$ with the x-axis. The supports of the flanking regions lie in the $xy$ plane at an angle of $\sim57^o$ degrees with the x-axis as shown in Fig. \ref{fig1}(b) and \ref{fig1}(c). This results in a 3d tetrahedral geometry where the imaginary continuations of all three posts meet at the d-cone, as shown schematically in Fig. \ref{fig1}(c) and \ref{fig1}(d). The sheet is rigidly clamped only to the central pole. The front conical section of the sheet is initially free and unsupported. The d-cone preparation and loading procedure is shown for a typical example in the supplementary movie S1.

Load is applied to the thin sheet in order to advance the d-cone by making use of two complimentary loading geometries: vertical and horizontal. Under both geometries two parallel cylindrical posts are fixed in the conical region of the d-cone at equal distances from the x-axis separated from each other by 25 mm,  and are displaced vertically at a constant rate of 0.5 mm/sec. In the vertical loading geometry the posts are perpendicular to the $xy$ plane; thus, only their tips push up against the sheet and the loading points remain fixed within the $xy$ plane of the lab frame of reference, as shown in Fig. \ref{fig1} (c.i). In the horizontal loading geometry the two posts are positioned parallel to the $xy$ plane so that the edge of the sheet slides along the loading posts as they are raised, therefore distancing the loading points form the d-cone, as shown in Fig. \ref{fig1} (c.ii). The loading posts are mounted on a Sartorius TE6101 scale, and the force that the sheet exerts against them is continuously recorded at a frequency of $3.8\pm0.1$ Hertz. The scale and the loading posts are placed on a Thorlabs LTS300 moving stage, and are displaced vertically as one unit. The deformation of the sheet is also imaged in real time by a digital camera (Thorlabs DCC1545M CMOS), positioned directly above it. We use photoelaticity to highlight the deformation process; the Mylar sheet is illuminated with a white LED from bellow and placed between two crossed linear polarizers. The Mylar sheets are anisotropic and display linear birefringence at zero strains. Therefore, in addition to the photo-elastic effect, the polarization of light transmitted through the Mylar is influenced by the relative orientation of the Mylar sheet with respect to the orientation of the polarizers, and with respect to the direction of light propagation.

\begin{figure}
\includegraphics[width=3.5 in,clip=] {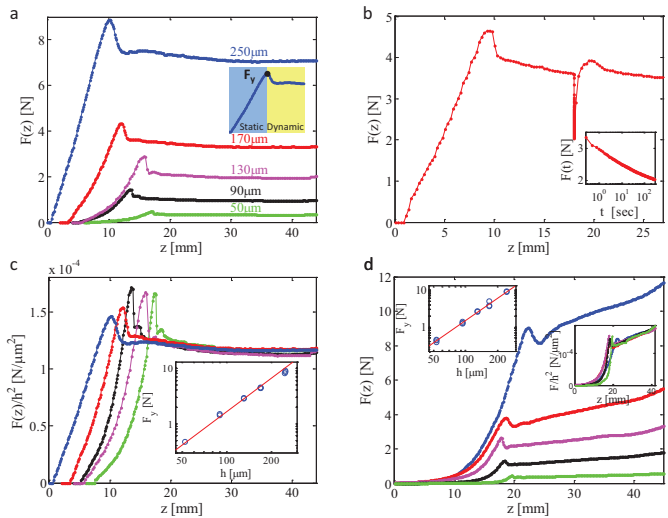}
\caption{\textbf{Force measurements.} (a) $F(z)$ for different $h$ under horizontal loading geometry. (inset) $F_y$ separates between the static and dynamic regime. (b) $F(z)$ for a 170 $\mu m$ thick sheet. At $z=18$ mm the loading posts are stopped and held in place for 300 seconds before increasing of $z$ is resumed. (inset) $F(h)$ as a function of time passed since stopping the loading posts. (c) $F(z)$ normalized by $h^2$. (inset) $F_y$ as a function of $h$. The red line of slope two is added as a guide to the eye. (d) $F(z)$ for vertical loading geometry. (insets) $F(z)$ normalized by $h^2$ and $F_y(h)$.}
\label{fig2}
\end{figure}

When the loading posts move up, the Mylar sheet resists its deformation and exerts a normal force, $F(z)$, against the posts. $F(z)$ has a characteristic functional form universal to all the different sheet thicknesses, $h$, as shown for a horizontal loading configuration in Fig. \ref{fig2}(a). We identify two distinct regimes in the evolution of $F(z)$: static and dynamic, highlighted by two color shadings in the inset to Fig. \ref{fig2}(a). In the static regime, the tip of the d-cone maintains its position throughout the loading process and $F(z)$ increases monotonically. When the force reaches a critical value, $F_y$, the d-cone yields, $F(z)$ abruptly decreases, and the d-cone propagates in the x-direction. We therefore refer to this regime as dynamic. Note, however, that if the loading is halted the d-cone arrests instantly, and $F(z)$ monotonically decays over time, as shown in the inset to Fig. \ref{fig2}(b). Once loading is resumed, a larger force is necessary for the d-cone to re-initiate motion, as shown in Fig. \ref{fig2}(b). The yield force required to initiate motion, $F_y$, scales as $h^2$, as shown in the insets to Fig. \ref{fig2}(c) and \ref{fig2}(d). The decrease in the force following $F_y$ is concomitant with the propagation of the d-cone and marks the transition from the static to the dynamic regime, as shown in the supplementary movie S2. In the dynamic regime, $F(z)$ remains nearly constant and, similarly to $F_y$, also scales as $h^2$. This is shown by the collapse of the curves following the yield event in Fig. \ref{fig2}(c) and inset to \ref{fig2}(d). Within the static regime $F(z)$ does not exhibit any simple scaling with $h$. Nevertheless, the rise of $F(z)$ to its critical value and the transition into the plastic regime occurs considerably more abruptly for thinner sheets.

\begin{figure}
\includegraphics[width=3.5 in,clip=]{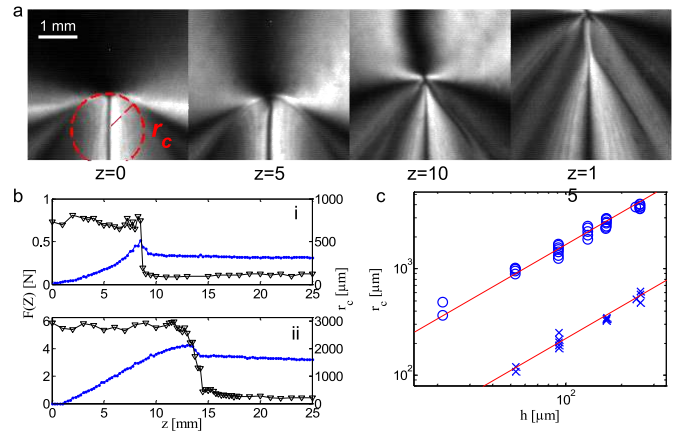}
\caption{\textbf{Propagation of the singular structure is preceded by the focusing of stresses at the tip.} (a) 4 typical snapshots showing a 50$\mu m$ thick sheet placed between two crossed linear polarizers at four different values of $z$. (b) $r_c(z)$ (triangles) and $F(z)$ (circles)  as a function of $z$ for  $h=50\mu m$ (i) and $h=170\mu m$ (ii) under horizontal loading geometry (c) $r_c$ vs. $h$ in the static regime (o) and dynamic regime (x). The two red lines of slope one are added as guides to the eye.}
\label{fig3}
\end{figure}

For the d-cone motion to ensue, the core structure of the d-cone must undergo a geometrical transition and focus the stresses at its leading edge. This is shown qualitatively in the sharpening of the d-cone between $z=5$ to $z=10$ in Fig. \ref{fig3}(a) and in the corresponding supplementary movie S2. In the static regime, the change in boundary conditions is accommodated by large-scale deformation over the length of the entire sheet while the shape and location of the d-cone remain practically unaffected, as shown for $z=0$ and $z=5$ in Fig. \ref{fig3}(a). Then, as the normal force reaches its critical value, the radius of the d-cone crescent-shaped core, $r_c$, dramatically decreases and the singularity starts to propagate with no further significant change in $r_c$. The abrupt focusing of $r_c$ occurs simultaneously with the drop of $F(z)$, as shown in Fig. \ref{fig3}(b) and in the supplementary movie S2. $r_c$ scales approximately linearly with the sheet thickness in both the static and dynamic regimes; however, within the dynamic regime the d-cone is approximately 7 fold sharper, as shown in Fig. \ref{fig3}(c). This linear scaling of $r_c$ is obtained over a range of only one decade and may display deviations from linearity if a wider regime will be accessible. However, it is impossible to reconcile our measurements with the scaling of $r_c\sim h^{1/3}$ obtained for d-cones in purely elastic sheets under normal load \cite{cerda2005confined,liang2005crescent}.

In the dynamic regime the loading obeys the scaling relation $F(z,h)=F(z)\cdot h^2$ for both loading geometries, as shown in Fig. \ref{fig2}(c) and the insets to Fig. \ref{fig2}(d).  This scaling can now be understood in terms of the linear dependence of deformation length scale, $r_c$, on the thickness of the sheet. We balance the work done by $F(z)$ with the dissipation associated with creating a furrow at the wake of the propagating d-cone. The d-cone focuses the deformation to a region of width $r_c\sim h$; giving rise to furrows of similar width. This irreversible deformation occurs throughout the thickness of the sheet, $h$; consequently, this results in a dissipation rate that is proportional to $h^2$, consistent with our measurements. This simple scaling argument suggests a propagation criterion for a d-cone that is analogous with the energy balance used to describe the crack propagation in fracture mechanics. However, this argument cannot account for the identical scaling exhibited by $F_y$. In this case the energy balance must also account for the focusing of the d-cone tip and the elastic energy associated with the abrupt drop in $F(z)$.

In the static regime, the d-cone is blunt and distributes the load over a wide region allowing the tip to persist under higher load before yielding. As the yield force is reached the d-cone tip sharpens, significantly focusing the applied load to a small region in which the plastic deformation takes place. The tip of the d-cone remains sharp throughout the dynamic stage localizing deformations more effectively; thus, requiring less force in order to set the d-cone in motion.

\begin{figure}
\includegraphics[width=3.5 in,clip=]{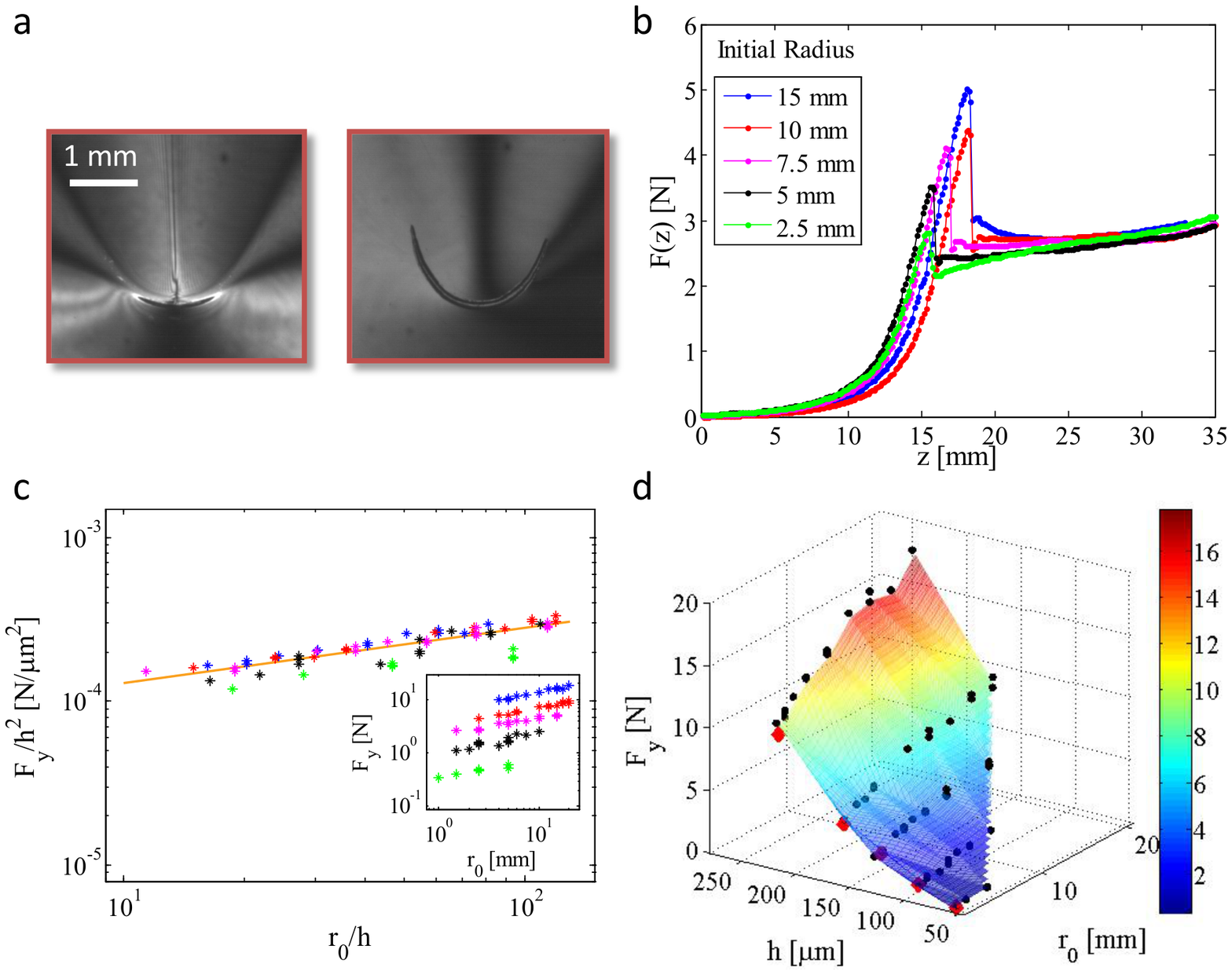}
\caption{\textbf{Imprinted crescent scars as d-cone surrogates.}  (a) Images of a Mylar sheet with a generic d-cone (left) and a laser-engraved crescent defect (right) mounted on the supporting posts and imaged between two crossed polarizers. (b) $F(z)$ for $h=130\mu m$ with varying initial radii, $r_0$. (c) $F_y/h^2$ as a function of $r_0/h$  for different $h$, color coded as in previous figures. The orange line of slope 1/3 is added as a guide to the eye. (inset) $F_y$ as a function of $r_0$ for different $h$. (d) 3d plot of $F_y$ as a function of $r_0$ and $h$. The surface $F_y(h,r_0)$ is interpolated numerically form the yield force values measured for the laser-engraved sheets (black dots). The red diamonds correspond to the average values of $F_y(h)$ measurement on generic d-cones.}
\label{fig4}
\end{figure}

The focusing of stresses at the tip of the d-cone is analogous to the parabolic displacement profile of a crack, where the singular profile of the crack effectively focusses elastic stress and all of the plastic dissipation is localized to a small processes-zone at the crack's tip. A common method to stop or delay crack propagation in brittle solids is to blunt their stress focusing mechanism. This is accomplished by rounding the otherwise sharp tip. We achieve the same goal for d-cones by using laser engraving to introduce well defined initial crescent shaped scars on the Mylar sheets. The crescent scars are weak lines for bending deformations and serve to guide the formation of the d-cones around a chosen core radius, $r_0$. This procedure does not alter the shape of the supported thin sheet away from the scar; nevertheless, in the vicinity of the scar the deformation of the core is significantly different than that of the generic d-cone, as shown in Fig. \ref{fig4}(a). We repeat the loading experiments with the laser-engraved Mylar sheets for radii $r_0$ ranging from 1 mm to 20 mm and find that the functional form of $F(z)$ in the static regime is insensitive to $r_0$; however, regions of higher initial curvature focus stresses more efficiently, resulting in higher values of $F_y$ for larger $r_0$ values, as shown in Fig. \ref{fig4}(b). The decrease of the force accompanying the onset of propagation is more abrupt for the artificial scars than for the pure d-cone. However, once the d-cone propagates away from the initial incision, $F(z)$ is again largely independent of $r_0$, and all curves converge to one. For a given sheet thickness, $F_y$ scales as $F_y\sim r_0^{1/3}$, as shown in the inset to Fig. \ref{fig4}(c). Additionally, when $F_y$ is rescaled by $h^2$ and plotted as a function of $r_0/h$ all the data collapses onto a universal master curve, as shown in Fig. \ref{fig4}(c); suggesting the following scaling for the critical force: $F_y(h,r_0)\sim h^{5/3}r_0^{1/3}$.

Despite the differences between the pure d-cone geometry and that around the artificial scar, the value of the yield force, $F_y(h,r_0)$, approaches the generic d-cone value, $F_y$, as $r_0$ approaches $r_c$, as indicated by the full red diamonds in Fig. \ref{fig4}(d). Values of $r_0$ below $r_c$ were experimentally inaccessible; bellow a critical radius the sheet cannot be manually forced into a d-cone like geometry around the smaller crescent scar. Interestingly, these minimal values of $r_0$ are very close to the initial $r_c$ values of the d-cone created by pushing a sheet through a ring. Yet, in spite of this apparent lower bound to the scale over which deformation can initially occur, it is surprising to note that after yielding the tip focuses further by nearly an order of magnitude. Further studies are required to identify the exact mechanism for the focusing of the d-cone and its scaling.

\bibliography{Bibiliography_PRL}

\end{document}